\documentclass{article}

\usepackage[preprint]{icbinb_neurips_2020}

\usepackage[utf8]{inputenc} 
\usepackage[T1]{fontenc}    
\usepackage{hyperref}       
\usepackage{url}            
\usepackage{booktabs}       
\usepackage{amsfonts}       
\usepackage{nicefrac}       
\usepackage{microtype}      
\usepackage{graphicx}
\usepackage{float}

\title{Improving non-deterministic uncertainty modelling in Industry 4.0 scheduling }

\author{%
  Ashwin Misra \\
  The Robotics Insitute\\
  Carnegie Mellon University\\
  Pittsburgh, PA \\
  \texttt{amisra@andrew.cmu.edu} \\
  \And
  Ankit Mittal \\
  Maruti Sizuki India \\
  Gurugram, India \\
  \texttt{anmittal01@gmail.com}\\
  \AND
  Vihaan Misra\\
  Netaji Subhas University of Technology\\
  New Delhi, India\\
  \texttt{vihaan.ee19@nsut.ac.in} \\
  \And
  Deepanshu Pandey\\
  ZS Associates\\
  Gurgaon, India\\
  \texttt{deepanshu.pandey@zs.com} \\
}

\begin{document}

\maketitle

\begin{abstract}
   The latest Industrial revolution has helped industries in achieving very high rates of productivity and efficiency. It has introduced data aggregation and cyber-physical systems to optimize planning and scheduling. Although, uncertainty in the environment and the imprecise nature of human operators are not accurately considered for into the decision making process. This leads to delays in consignments and imprecise budget estimations. This widespread practice in the industrial models is flawed and requires rectification. Various other articles have approached to solve this problem through stochastic or fuzzy set model methods. This paper presents a comprehensive method to logically and realistically quantify the non-deterministic uncertainty through probabilistic uncertainty modelling. This method is applicable on virtually all Industrial data sets, as the model is self adjusting and uses epsilon-contamination to cater to limited or incomplete data sets. The results are numerically validated through an Industrial data set in Flanders, Belgium. The data driven results achieved through this robust scheduling method illustrate the improvement in performance.
\end{abstract}

\section{Introduction}
Despite the onset of Industry 4.0 across various Manufacturing plants, many areas of this relatively newer framework needs improvement to enhance it's reliability and accuracy. Production planning combines and coordinates all the manufacturing activities; it broadly consists of three components- planning, controlling and dispatching. The planning process refers to the pre-manufacturing task of the objectives and targets with respect to the available resources and constraints. Control refers to the continuous evaluation of the performance, according to the standards set by planning. Hence, the aim of production control is to make sure that the consignment is produced at the optimum quality, time and quantity with cost-effective methods. In Production Planning, Scheduling refers to the part of planning which is concerned with the schedule of an activity; the start time and the finish time. This paper aims to develop a robust scheduling strategy to improve the scheduling aspect of production planning.  

The main reason, the scheduling of a consignment is inconsistent or deviates from the standard behaviour is due to various non-deterministic reasons such as human operator error, machine faults, delays in supply etc. The current planning algorithms do not capture these intrinsic non-deterministic uncertainties which result in poor performance during execution. This leads to losses in capital, time and resources. The method discussed in this paper is influenced by this problem. These uncertainties are more accurately quantified to develop an improved robust scheduling strategy.    

This paper provides a general outline to more realistically quantify operational non-deterministic uncertainty which is conventionally different from the general practices currently employed in the industry.The general practices assume it to be a deterministic system, which leads to such faulty predictions[1]. It introduces the uncertainty models used- P-box and epsilon-contamination models, briefly describe the Industrial data set and explain how to model on such data. A numerical example is introduced for a specific case which would enhance the understandability of the project.

\subsection{Current Literature}

During the last years, a lot of academic research has focused on the topic of production planning and scheduling. The authors [2] attribute the operational uncertainty to software that are fixated on ideal schedules. They suggest human operator intervention from the production planner to the task scheduler- depending on the frequency and severity of scheduling errors. These respondents are responsible for rescheduling and identifying the optimal path of action. This method relies on human intervention to counter the planning faults due to non-deterministic uncertainty. However, this is only applicable to small production plants and does not factor in the human operator errors.

The authors[3] have focused on reviewing methods to scheduling under uncertainty including robust, stochastic and fuzzy scheduling. They propose a baseline schedule which is planned before the manufacturing operation. Reactive scheduling re-optimizes this schedule dynamically according to the uncertainties. Hence it depends on a stochastic assessment of the uncertainty, from which a set of decisions are developed.

\section{Uncertainty Modelling}
Generally, in such systems, the uncertainty is either quantified as probabilistic[4] or fuzzy set models[5]. Typically, stochastic methods are used with a combination of different probability distribution models to represent various phenomenons. 
These models are however unrealistic due to the fact that interval-based uncertainties are not possible to quantify/model with fixed parameter probability distributions. They fail to take into account the variations in uncertainty such as the variation in any initial probability distributions such as median value shifts. Such models are also very complex to build and require field experts to define and generate. 
As shown by [5], classic probabilistic models fail to capture the whole non-deterministic nature of uncertainties in manufacturing processes. After analysis of a large Industrial data set, it was concluded that probability box models are much more accurate for manufacturing plants.
Epsilon contamination is a method to capture uncertainty through e-contamination classes. It is a bayesian method, which is used in data sets which has corrupted, incomplete or insufficient data. It is used in Robust statistics and the methodology discusses uses this as one of its components.

\subsection{Probability box}
A P-box or a probability box is the area circumscribed within an upper and lower cumulative distribution function to characterize an imprecise distribution. The most significant benefit of a p-box is that it captures this non-deterministic uncertainty within a confined group very well in juxtaposition to Probability distribution functions themselves. For a more extensive data set, set are divided into a subset, and PDFs are computed, then PDFs are integrated to construct a bounded set of CDF. The non-deterministic likelihood of the technique or parameters affected by this new unpredictable change is represented by this set of CDFs. 

\begin{equation}
 F(x) = P(X <= x)
\end{equation}
A Cumulative Frequency distribution represents the cumulative probability of a random variable from negative infinity up to a random variable X. By definition, F is a non-decreasing function with a range of $[0,1]$ over the domain of $[-\infty, \infty]$
\begin{equation}
 f(x) = Pr[X = x]
\end{equation}
\begin{figure}[H]
  \centering
  \fbox{\includegraphics[scale=0.035]{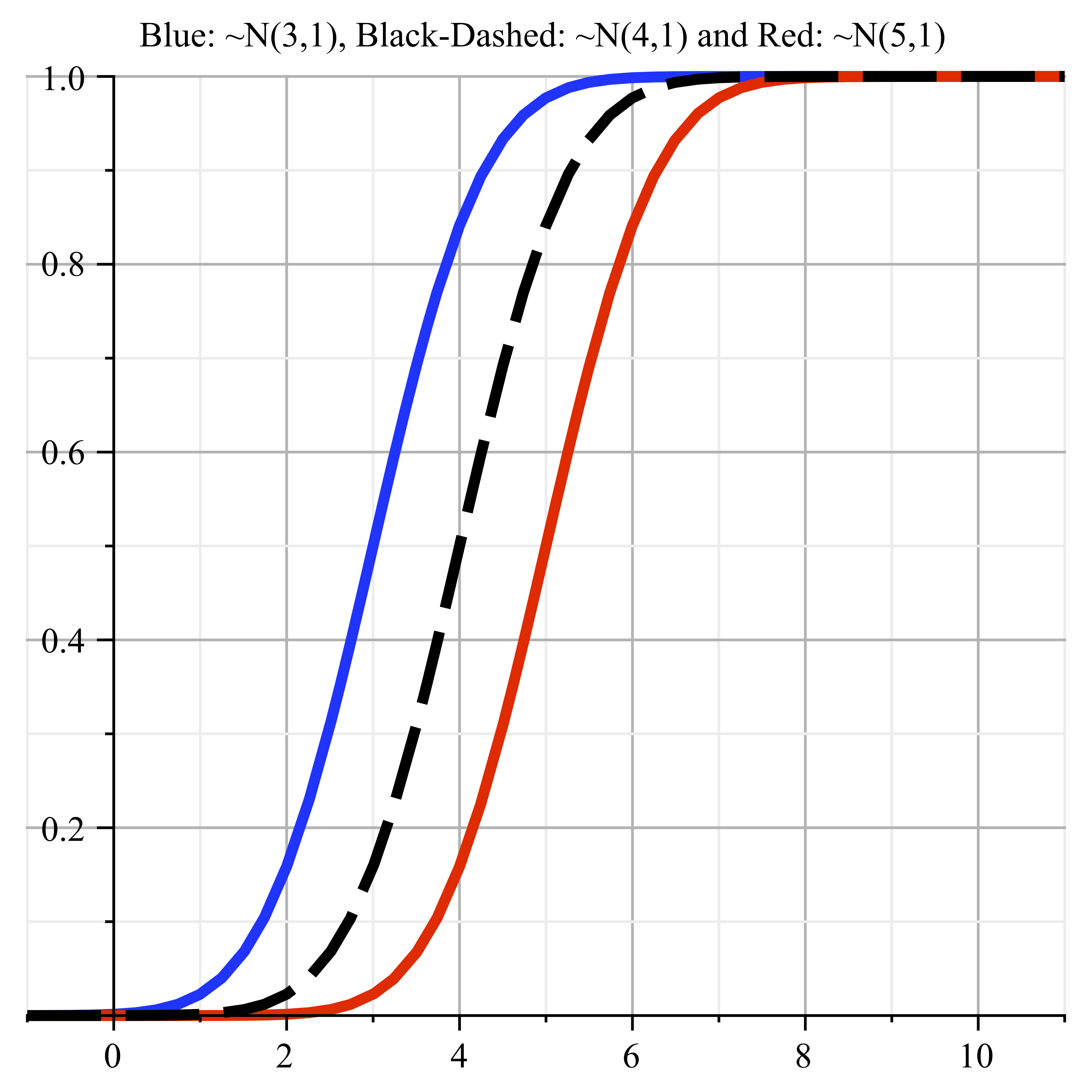} \rule[-.5cm]{0cm}{0cm} \rule[-.5cm]{0cm}{0cm}}
  \caption{Sample p-box}
\end{figure}
The PDF is the derivative of the CDF if it exists. A generalised probability box, or generalised p-box, is a pair $(\underline{F},\overline{F})$ of distribution functions from $\Omega$ to [0, 1] following $F<=F$. If $\Omega$ is a closed interval on R, then we call $(\underline{F},\overline{F})$ a p-box. When both $\underline{F},\overline{F}$ assume general number of different values only, It can be said that the generalised p-box is discrete. $\underline{F}_{X}(x)$ and $\overline{F}_{X}(x)$ are respectively the lower and upper cumulative probability bounds, and ${F}_x(x)$ is non-decreasing with x,
$\underline{F}_{X}(x) = \underline{P}(X<= x)$,  $\overline{F}_{X}(x) = \overline{P}(X<= x)$

As a function of unexplained unpredictable disturbances to specific techniques or parameters, p-boxes are a valuable method for catching non-deterministic uncertainties. Let's consider a particular task performed by 2 workers, and the time taken to complete the job is recorded for a year. Using the data, respective PDFs and CDFs are plotted. Then from the resulting CDFs, it is computed that which worker works better than others for the particular task. But the result did not account for the non-deterministic uncertainty parameters such as mood or the worker's distractions. It may be possible that these uncertainties affect the worker timing and due to which one CDF has a higher variation than others. To capture these non-deterministic uncertainties, the p-box is the most suitable tool. To plot the p-box data set is divided into subsets; upper and lower bounded CDFs are plotted. P-box captures the uncertainty in the probability of duration task completion. This enables the system for more robust task assignments for any future event accommodating non-deterministic delays.

\subsection{Epsilon Contamination}
The epsilon contamination model
\begin{equation}
 \underline{P}(f)=(1-\epsilon)P(f)+ \epsilon \underline{Q}(f)
\end{equation}
In this model[7], the probability epsilon ($\epsilon$) donates the probability that the data set is contaminated by a distribution Q. Let P be the distribution of the sequence i.e. a function of the measures of central tendency. This model provided by Huber proposes a robust statistical framework by establishing priors. It is a convex combination of two distinct uncertainty models- 1. Vacuous Model and 2. Probabilistic model.
Epsilon, is a parameter assigned to the trust on the model. 

Both of these models rely on the principle of imprecise belief by [6], the intervals of the p-box and epsilon contamination models quantify uncertainty accurately and should be incorporated in the scheduling algorithms. This belief can be attributed to an operators experience and his expertise in a particular task, i.e. assigning a particular factor to replicate the belief of the process. In this paper, the specific distributions used are cumulative distribution functions.

\section{Modelling on Industrial Data set}
This section will explain how to model a particular industrial process through the uncertainty models and how to interpret the model to achieve a more accurate scheduling and robust planning autonomous framework. The modelling technique is divided into two parts- For data with limited information and for complete data. In this numerical case, a critical point of n=25 samples is taken. For the incomplete data set, the epsilon-contamination model is applied with a belief measure. P-boxes are applied on the complete data sets as they have restricted bounds due to the complete information. 
\\
The database obtained from the industry needs to be pre-processed before entering into this tool if it is unstructured. As the main objective of this research is to optimize the scheduling. In our example, The attributes of the database which exist for operation time, operator ID, task Sequence and Operator skill are taken. Any similar attributes pointing to similar sets of attributes should be identified and taken according to different Industries.
\\
For each task undertook in the industry, various operational parameters are considered. Every sequence, operator and season corresponding to a particular task.
For every season, cumulative distribution functions of every operator executing a task is plotted. 
\\
The error times are calculated for every operator in the different operations and season timings. 
\begin{equation}
    \varepsilon= Predicted Time- Observed Time
\end{equation}

This error is equal to the Actual time recorded minus the predicted time by the IoT-prediction software. Cumulative distribution functions and probability density functions are plotted for each.
After analyzing the graphs of the cumulative distribution functions for very operator per operation sequence. Now the curves are of two types-\\
First, In which the data points are sufficient to plot probability box for the operator, Second, Less than sufficient data points to form a probability box. In the second case, the epsilon contamination model proves to be a useful analytical tool.
\\ In the first case, with sufficient data points, the uncertainty is quantified with p-box models as shown in the figure. The numerical example of SeqID786 shows the CDF of various operators executing the same operation, The maximum and minimum bounds of this figure gives the upper and lower bounds/previsions of the p-box. This figure denotes the uncertainty of a specific operator using upper and lower previsions. The area bound between the upper and lower previsions denotes the uncertainty or error in the task execution. Similar graphs are plotted for all operators to study their error patterns across various seasons, tasks and operation parameters. The upper bound of this curve denotes the upper prevision and the lower bound denotes the lower prevision.
\begin{figure}[H]
  \centering
  \fbox{\includegraphics[scale=0.5]{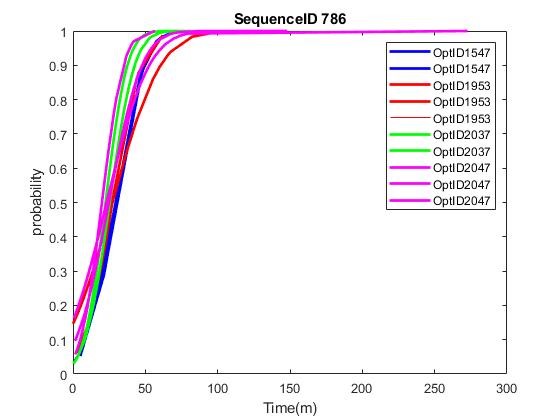} \rule[-.5cm]{0cm}{0cm} \rule[-.5cm]{0cm}{0cm}}
  \caption{p-box of sample sequenceID 786}
\end{figure}
For the second case, for lesser data points or the operators who are on a temporary basis or have limited responsibilities. Epsilon contamination is a very new research topic which focuses on a small region and generalizes for the whole range. The trust factor is taken as 0.8. In the given example, seqID 787 is taken which had only three operators working on this sequence and only 9 observations. To capture the non-determinism in this data, epsilon contamination samples the distribution to form upper and lower previsions according to a trust factor $\epsilon$.

The upper and lower previsions are calculated using this formula and their difference denotes the degree of uncertainty in the data. In this formula, the trust factor is denoted by epsilon, which is assigned by an Industry expert/supervisor on the probability of the task being completed in time. The contaminated model means the lower and upper previsions are calculated from the average distribution of the operators in a sequence with an individual distribution contaminant. 
\begin{figure}[H]
  \centering
  \fbox{\includegraphics[scale=0.2]{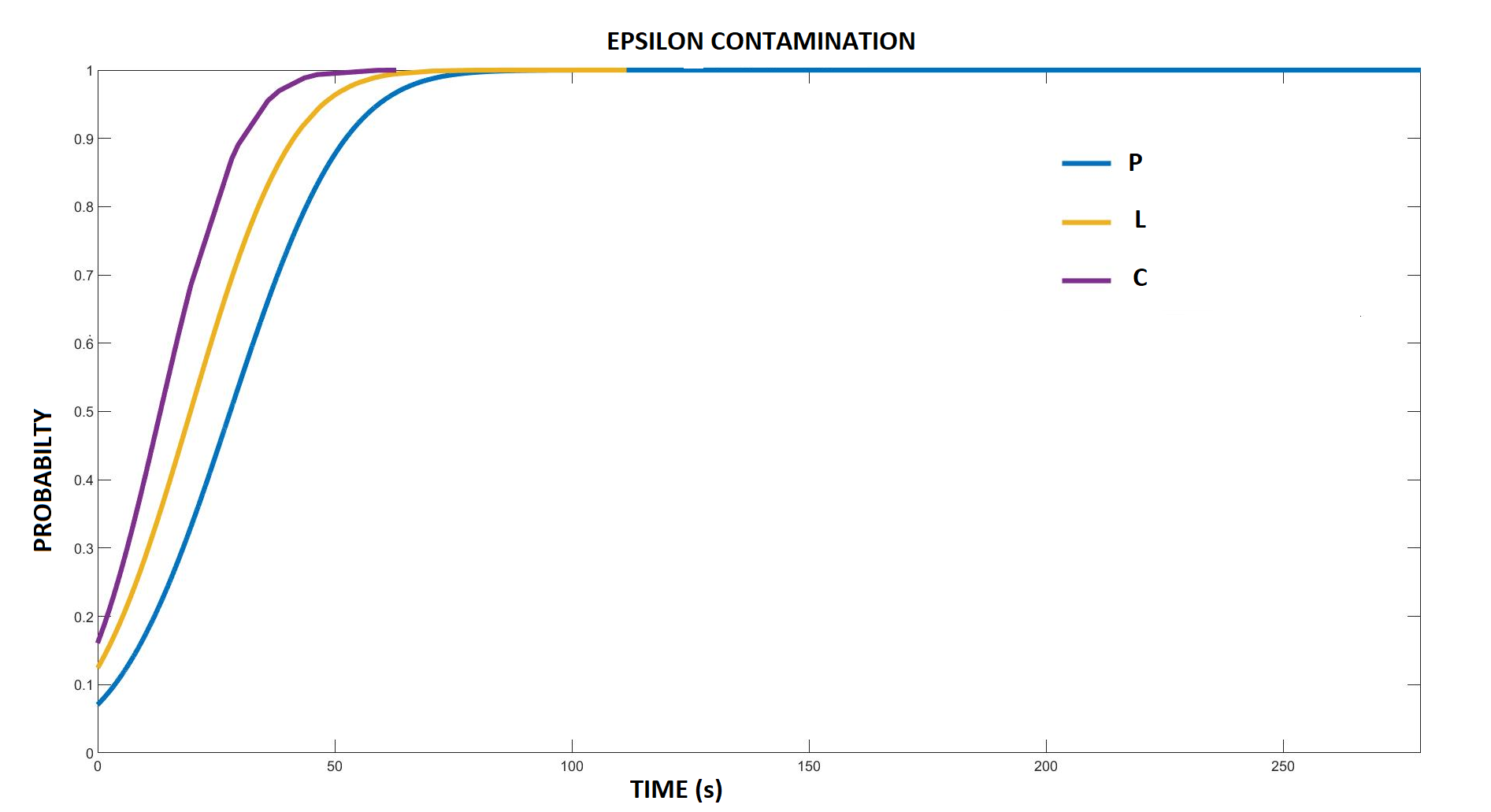} \rule[-.5cm]{0cm}{0cm} \rule[-.5cm]{0cm}{0cm}}
  \caption{Epsilon Contamination model, P=Upper Prevision, C=Lower Prevision, L=Mean prevision curve}
\end{figure}
The Second step in the process after quantifying the uncertainty is to improve the prediction model of the Industry 4.0 framework. The current task completion time prediction used is not accurate and differs greatly from the real time recorded. This model can be improved using dynamic Machine Learning;
\begin{equation}
    \overline{\varepsilon} = f(estimated production time, training data)
\end{equation}
And the predicted estimated time according to the new model is
\begin{equation}
    New estimated time = f+\overline{\varepsilon}
\end{equation}
To give an insight, in the paper [3], a Rational Quadratic Gaussian method Regression Model was chosen when a lot of models were analyzed on this knowledge, the model was trained stochastically on the past estimated and real production times to estimate the assembly time with decreased error. The training data is used for an year of operation with the exception of operators who have very less data points. The model is explained as follows-
\begin{equation}
     \overline{\varepsilon}= N(f(x, b), a^2)
\end{equation}
The model has to be suited to different data sets. The numerical example have employed this with parameters $b = 7.734x10^3$ with $a = 4.43x10^3$. The result section compares the recomputed p-box which shows that this learning model improved the prediction times are reduced the error times by an appreciable margin.
\\ The final step is consolidating all these features into an easy-to-use GUI. This will increase the usability of the research results and will lead to wider adaptability to different industries. An application can be developed via MATLAB or python to incorporate all such modelling techniques suggested in this paper and link it with a regression learner to automate the whole process. The best operator suggestion is made by calculating the degree of uncertainty i.e. the difference of the upper and lower previsions. The predicted time of completion of operation is also shown using the new developed model.
\\
\section{Results}
\begin{figure}[H]
  \centering
  \fbox{\includegraphics[scale=0.5]{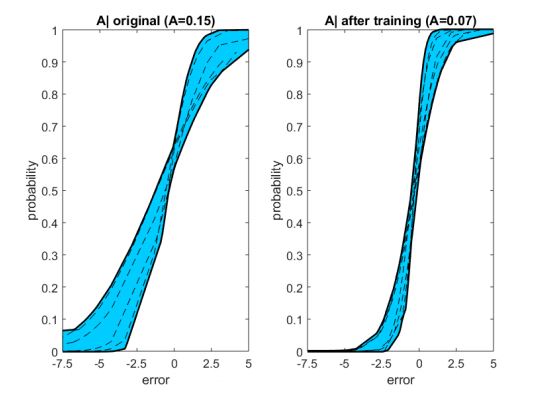} \rule[-.5cm]{0cm}{0cm} \rule[-.5cm]{0cm}{0cm}}
  \caption{P-box comparison, before and after training}
\end{figure}
The results achieved through the adoption of this process by the facility are very substantial. The industry has reported more accurate prediction times and have helped in estimating the delivery time for different operations.
After, the machine learning process the P-boxes are again computed which highlight that the error times are reduced. A reduction is e.g., observed from Original = $0.15$ units to TrainedModel = $0.7$ for the same operation as per normalised area. As the area under the pox-describe the variance in the error values, this shows that the predicted times are now closer to the actual times observed. This research has also helped to rank the employees on their performance and create a new assignment schedule for different operations and operators according to different seasons.

\section{Conclusion}
To conclude, this article provides a new perspective to data optimization of Industry 4.0. This article bridges the gap between pure mathematics and application and proposes a novel applied mathematics approach to quantify non-deterministic uncertainty. General practices assume determinism which is an inherent flaw in the procedure and this paper proposes a dynamic approach to solve this problem.

\section*{References}

\small
[1] Hendrik, V.B., \& Paul, V., (2017). “Design of holonic manufacturing systems,” Journal of Machine Engineering, vol. 17/3.

[2] Snoo, C., Wezel, W., Wortmann, H., \& Gaalman, G.
(2011). Coordination activities of human planners during rescheduling: Case analysis and event handling procedure.
International Journal of Production Research,
49, 2101–2122. doi:10.1080/00207541003639626

[3] Herroelen, W. \& R. Leus,  (2005). “Project scheduling under uncertainty: Survey and research potentials.” Eur. J. Oper. Res. 165: 289-306.

[4]Knyazeva, M., Bozhenyuk, A., \& Rozenberg, I., (2015).
Resource-constrained project scheduling approach
under fuzzy conditions. Procedia Computer Science, 77,
56–64. doi:10.1016/j.procs.2015.12.359

[5] Shariatmadar, K., Misra, A., Debrouwere, F., \& Versteyhe,
M. (2019). Optimal modelling of process variations in
industry 4.0 facility under advanced p-box uncertainty.
IEEE Student Conference on Research and
Development (SCOReD), Bandar Seri Iskandar,
Malaysia, 2019, pp. 180–185

[6] Walley, P. (1991). Statistical reasoning with imprecise probabilities. Taylor \& Francis.

[7]Chen, M., Gao, C. \& Ren, Z., (2015). A General Decision Theory for Huber's $\epsilon$-Contamination Model. Electronic Journal of Statistics. 10. 10.1214/16-EJS1216.

\end{document}